\documentclass[twoside]{llncs}
\usepackage{amsmath}
\usepackage{amssymb}
\usepackage{graphics}
\usepackage[dvips]{epsfig}
\usepackage{times}
\usepackage{algorithm}
\usepackage{algorithmic}
\usepackage{fancyhdr}
\usepackage{ifthen}
\newcommand\settitle[2][]{%
 \title{#2}
 \ifthenelse{\equal{#1}{}}%
  {\fancyhead[RO]{\nouppercase #2 \qquad \thepage}}%
  {\fancyhead[RO]{\nouppercase #1 \qquad \thepage}}%
}
\newcommand\setauthors[2]{%
 \author{#2}
  {\fancyhead[LE]{\thepage \qquad \nouppercase #1}}%
}
\fancypagestyle{plain}{%
\fancyhf{}

}

\def\keywordsname{Keywords.}
\newenvironment{keywords}{%
      \list{}{\advance\topsep by-0.50cm\relax\small
     \leftmargin=1cm
      \labelwidth=1cm%\z@
     \listparindent=1cm%\z@
     \itemindent\listparindent
      \rightmargin\leftmargin}\item[\hskip\labelsep
                                    \bfseries\keywordsname]}

   {\endlist}

   \newtheorem{Alg}{Algorithm}

   \def\C{{\mathbb C}}
   \def\R{{\mathbb R}}
   \def\N{{\mathbb N}}

   \def\S{{\Sigma}}

   \def\Pf{{\it Proof.$\;\;$}}

   \def\qed{\hfill$\diamond$}
   
   \def\rk{\mbox{\rm rk~}}
   \def\spann{\mbox{\rm span}}
   \def\tr{\mbox{\rm tr~}}

   \def\cA{{\mathcal A}}
   \def\cM{{\mathcal M}}
   \def\cQ{{\mathcal Q}}
   \def\cS{{\mathcal S}}
   
   \def\cP{{\mathcal P}}

   \def\la{\leftarrow}

\def\Z{{\mathbb Z}}
\def\cC{{\mathcal C}}
\def\cR{{\mathcal R}}
\def\ve{{\bf e}}
\def\vf{{\bf f}}
\def\vg{{\bf g}}
\def\vp{{\bf p}}
\def\vE{{\bf E}}
\def\vF{{\bf F}}
\def\ora{\overrightarrow}
\def\ola{\overleftarrow}
\begin{document}

\settitle[Efficient Equivalence Tests]{Efficient Tests for Equivalence
  of Hidden Markov Processes and Quantum Random Walks}

\setauthors{U.Faigle/A. Sch\"onhuth}
           {Ulrich Faigle$^1$ and Alexander Sch\"onhuth$^2$, Member IEEE}
\institute{Mathematical Institute\\
           Center for Applied Computer Science\\
           University of Cologne\\
           Weyertal 80\\
           50931 K\"oln, Germany\\ [1ex]
%\email{faigle@zpr.uni-koeln.de}\\[1ex]
\and
           Centrum Wiskunde \& Informatica\\
           Science Park 123\\
           1098 XG Amsterdam\\
\email{as@cwi.nl}}

%------------------------------------------------------------------------
\date{}
\maketitle

\thispagestyle{plain}
%--------------------------------------------------------------------------
\begin{abstract}
\begin{sloppypar}
While two hidden Markov process (HMP) resp.~quantum random walk (QRW)
parametrizations can differ from one another, the stochastic processes
arising from them can be equivalent. Here a polynomial-time algorithm
is presented which can determine equivalence of two HMP
parametrizations $\cM_1,\cM_2$ resp.~two QRW parametrizations
$\cQ_1,\cQ_2$ in time $O(|\S|\max(N_1,N_2)^{4})$, where $N_1,N_2$ are
the number of hidden states in $\cM_1,\cM_2$ resp.~the dimension of
the state spaces associated with $\cQ_1,\cQ_2$, and $\S$ is the set of
output symbols.  Previously available algorithms for testing
equivalence of HMPs were exponential in the number of hidden
states. In case of QRWs, algorithms for testing equivalence had not
yet been presented.  The core subroutines of this algorithm can also
be used to efficiently test hidden Markov processes and quantum random
walks for ergodicity.
\end{sloppypar}
%of for the equivalence-testing problem (ETP)
%for hidden Markov models (HMMs), based on a novel algebraic theory for
%random sources, is presented. Our algorithm two HMMs $\cM_1$ and $\cM_2$ It gives rise to an efficient and
%practical algorithm that can be easily implemented. Extant approaches
%are exponential in the number of hidden states and therefore only
%applicable to a limited degree. The algorithm can be equally applied
%to solve the IP for quantum random walks (QRWs) that have recently
%been presented as an analogon of Markov chains in quantum information
%theory. Moreover, the algorithm can be used to efficiently test HMPs
%and QRWs for ergodicity, which had remained an open problem so far.
\end{abstract}

%------------------------------------------------------------------------
%------------------------------------------------------------------------

\section{Introduction}
\label{sec.introduction}

Let a parameterized class of stochastic processes be described by a
mapping
\begin{equation}
\Phi:\cP\to\cS
\end{equation}
where $\cP$ is the set of parameterizations and $\cS$ is the corresponding
set of stochastic processes. 
A stochastic process $\Phi(P)$ as induced by the parameterization $P$ is 
said to be {\em identifiable} iff
\begin{equation}
\Phi^{-1}(\Phi(P)) = \{P\}
\end{equation}
that is, iff the parameterization giving rise to it is uniquely
determined.  The entire class of stochastic processes $\Phi(\cP)$ is
said to be {\em identifiable} iff $\Phi:\cP\to\Phi(\cP)$ is
one-to-one.  The \emph{equivalence problem} (EP) emerges when $\Phi$
is many-to-one and is to decide whether two parameterizations
$P_1,P_2$ are equivalent, that is $\Phi(P_1)=\Phi(P_2)$. Understanding
its solutions can significantly foster understanding of the classes of
stochastic processes under consideration as it usually yields insights
about the class' complexity and its number of free
parameters. Therefore, apart from its theoretical relevance, it is an
important issue in the practice of system identification
(e.g.~\cite{Pintelon01}). \par \begin{sloppypar} \emph{Hidden Markov
    processes (HMPs)} are a class of processes which have gained
  widespread attention. In practical applications, for example, they
  have established gold standards in speech recognition and certain
  areas of computational biology. See
  e.g.~\cite{Rabiner89,Durbin,Ephraim02} for comprehensive related
  literature. In an intuitive description, a hidden Markov process is
  governed by a Markov process which, however, cannot be observed.
  Observed symbols are emitted according to another set of
  distributions which govern the hidden, non-observed states. Since
  observed processes can coincide although the non-observed processes
  on the hidden states can differ from one another, hidden Markov
  processes are non-identifiable.\par\end{sloppypar} For hidden Markov
  processes, the EP was first discussed in 1957 \cite{Blackwell57}
  (see also \cite{Gilbert59} for a subsequent contribution). It was
  formulated for finite functions of Markov chains (FFMCs), an
  alternative way of parametrizing hidden Markov processes where, as
  sets of parametrizations, the parametrizations discussed here, also
  referred to as {\em hidden Markov models (HMMs)} in the following,
  models trivially contain FFMCs. The EP for hidden Markov processes
  was fully solved in 1992 \cite{Ito92}. The corresponding algorithm
  is exponential in the number of hidden states and therefore
  impractical for larger models. See \cite{Ito92} also for more
  related work.\par {\em Quantum random walks (QRWs)} have been
  introduced to quantum information theory as an analog of classical
  Markov sources \cite{Aharonov01}.  For example, they allow emulation
  of Markov Chain Monte Carlo approaches on quantum computers. A
  collection of results has pointed out that they would be superior to
  their classical counterparts with respect to a variety of aspects
  (see e.g.~\cite{Kempe05,Ambainis05,Marquezino08}). However, although
  their mechanisms can be described in terms of elementary linear
  algebraic definitions, their properties are much less
  understood. The key element of a quantum random walk parametrization
  is a graph whose vertices are the observed symbols. Quantum
  probability distributions on the vertices are transformed by linear
  operations which describe the quantum mechanical concepts of
  evolution and measurement. It is easy to see that quantum random
  walks are non-identifiable. For example, any of the (infinitely many
  different) parametrizations with a graph of only one vertex yields
  the same, trivial process. The equivalence problem for quantum
  random walks has not been discussed before.\par Beyond the work
  cited in \cite{Ito92}, there is a polynomial-time solution to test
  equivalence of probabilistic automata \cite{Tzeng92} where HMMs can
  be viewed as probabilistic automata with no final probabilities
  \cite{Dupont05}.  The crucial difference, however, is that
  probabilistic automata do not give rise to stochastic processes
  (distributions over infinite-length sequences), but to probability
  distributions over the set of strings of finite length.  The
  algorithm presented in \cite{Tzeng92} decisively depends on this and
  therefore does neither apply for hidden Markov processes nor for
  quantum random walks. Conversely, by adding a stop symbol to the set
  of observed symbols, any probability distribution over the set of
  strings of finite length can be viewed as a probability distribution
  over the set of infinite-length symbol sequences. This way, it can
  be seen that our solution also applies for probabilistic automata
  and therefore is more general than \cite{Tzeng92}'s solution.\\

 Overall, the purpose of this work is to present a simple,
 polynomial-time algorithm that solves the EP for both hidden Markov
 processes and quantum random walks:

\begin{theorem}
\label{t.mainresult}
Let $\S$ be a finite set of symbols and
\begin{equation}
\cM_X,\cM_Y\quad\text{resp.~}\quad\cQ_X,\cQ_Y
\end{equation}
be two hidden Markov process resp.~quantum random walk parametrizations
giving rise to the processes $(X_t),(Y_t)$ emitting symbols
from $\S$. Let
%\begin{equation}
%n:=\max\{n_X,n_Y\}
%\end{equation}
\begin{equation}
n_X,n_Y
\end{equation}
be the cardinalities of the set of hidden states in the hidden Markov
models resp.~the dimensions of the state spaces associated with the
quantum random walks. Equivalence of $(X_t),(Y_t)$ can be determined
in
\begin{equation}
O(|\S|\max\{n_X,n_Y\}^4)
\end{equation} 
arithmetic operations.
\end{theorem}

\begin{remark}
\begin{sloppypar}
Note that a polynomial-time solution for the identifiability problem
for HMPs does not provide a polynomial-time solution for the graph
isomorphism problem. There are both non-equivalent HMPs which act on
sets of hidden states which are isomorphic as graphs (e.g.~two HMPs
both acting on only one hidden state which, however, have different
emission probability distributions) and equivalent HMPs where
underlying graphs are non-ismorphic (e.g.~two HMPs, one acting on two
hidden states, but emitting the symbol $a$ with probability $1$ from
both states and the other one acting on only one hidden state, also
emitting the symbol $a$ with probability $1$, both result in the
stochastic process which generates $aaaa....$ with probability $1$).
\end{sloppypar}
\end{remark}

\begin{remark}
In \cite{Schoenhuth07} it was described how to test HMPs for
ergodicity. Plugging the algorithm for computation of a basis
(see subsection \ref{ssec.basis-computation}) into the generic
ergodicity tests provided in \cite{Schoenhuth07} renders these tests
efficient.
\end{remark}

\subsection{Organization of Sections}

The core ideas of this work are tightly interconnected with the theory
of \emph{finitary processes}. Therefore, we start by concisely
revisiting their theory in section \ref{sec.finitary}.  We then
introduce hidden Markov models and quantum random walk
parametrizations and the mechanisms which give rise to the associated
processes in section \ref{sec.equivalence}. In section
\ref{sec.computations} we outline how to most efficiently compute
probabilities for both hidden Markov processes and quantum random
walks. The algorithm and theorems behind our efficient equivalence
tests are then presented in section \ref{sec.equivalencetests}. We
finally outline some complementary applications of our algorithms and
make some conclusive remarks.

\section{Finitary Random Processes}
\label{sec.finitary}

Throughout this paper, we consider discrete random processes $(X_t)$
that take values in the (fixed) finite alphabet $\Sigma$. We assume
that the process emits the \emph{empty word} $\Box$ at time $t=0$. We
denote the \emph{probability function} $p$ of $(X_t)$
by 
\begin{equation} p(a_1\ldots a_t) :=\Pr\{X_1 = a_1,\ldots,
  X_t=a_t\} \quad(a_1\ldots a_t\in \Sigma^t).
\end{equation}
As usual, we set
\begin{equation}
     \Sigma^* :=\bigcup_{t\geq 0} \Sigma^t \quad\mbox{(with $\Sigma^0 =\{\Box\}$)}
\end{equation}
and note that $\Sigma^*$ is a semigroup under the concatenation $wv\in
\Sigma^{s+t}$ for $w\in \Sigma^s$ and $v\in \Sigma^t$. $|v| = \ell$ is
the \emph{length} of a word $v\in \Sigma^\ell$.

\medskip
For any $v,w\in \Sigma^*$, we define functions $p_v, p^w: \Sigma^*\to \R$ {\it via}
\begin{equation}
        p_v(w) := p(wv) =: p^w(v)
\end{equation}
and view $\R^{\Sigma^*}$ as a vector
space. $p(v=v_1...v_t|w=w_1...w_s)$ generally denotes the
\emph{conditional probability}
\begin{equation}
\begin{split}
   p(v|w) &:= \Pr(X_{s+1}=v_1,...,X_{t+s}=v_t\,|\,X_1=w_1,...,X_s=w_s)\\ 
   &= \left\{\begin{array}{cl} p(v) &\mbox{if $w=\Box$}\\
               0 & \mbox{if $p(w) = 0$}\\
               p(wv)/p(w) &\mbox{otherwise.}\end{array}\right.
\end{split}
\end{equation}

Furthermore, the subspace
\begin{equation}
\cR(p) := \spann\{p_v\mid v\in \Sigma^*\}
\quad\mbox{resp.}\quad  
\cC(p) :=\spann\{p^w\mid w\in \Sigma^*\}
\end{equation}
is the \emph{row space} resp. \emph{column space} associated with the
(probability function $p$ of) the random process $(X_t)$.

\medskip
It is easy to see that $\cR(p)$ and $\cC(p)$ have the same vector
space dimension. So we define the \emph{dimension} of $(X_t)$ (or its
probability function $p$) as the parameter
\begin{equation}
    \dim(X_t) = \dim(p) \;:=\; \dim\cR(p) = \dim\cC(p) \;\in \Z_+\cup\{\infty\}.
\end{equation}
For any $I,J\subseteq \Sigma^*$, we define the matrix
\begin{equation}
     P_{IJ} :=  [p(wv)]_{v\in I,w\in J} \in \R^{I\times J}.
\end{equation}
$P_{IJ}$ is called \emph{generating} if $\rk(P_{IJ}) =
\rk(P_{\S^*,\S^*}) (=\dim(p))$ and \emph{basic} if it is generating
and minimal among the generating $P_{IJ}$, that is $|I| = |J|=\dim(p)$
in case of $\dim(p)<\infty$. In that sense, we call (in slight abuse
of language) $I$ resp.~$J$ a \emph{row} resp.~\emph{column
  generator/basis} and the pair $(IJ)$ a \emph{generator/basis} for
$p$.\\

We call a process $(X_t)$ \emph{finitary} if it admits a (finite)
basis. So the finitary processes are exactly the ones with finite
dimension.

\medskip
\begin{remark}
The dimension of a random process is known as its {\em minimum degree
  of freedom}. The term \emph{finitary} was introduced in
\cite{Heller65}. Finitary processes are also called {\em linearly
  dependent} \cite{Jaeger00}.
\end{remark}

\medskip
\begin{theorem}
\label{t.key-theorem} 
Let $(X_t)$ and $(Y_t)$ be discrete, finitary random processes (over
$\Sigma$) with probability functions $p$ and $q$. Let furthermore
$(IJ)$ be a basis for $(X_t)$. Then the following statements are
equivalent:
\begin{itemize}
\item[(a)] $p = q$.
\item[(b)] $(I,J)$ is a basis for $(Y_t)$ and the equalities 
  \begin{equation}
    \label{eq.keyequalities}
    p(v) = q(v),\qquad p(wv) = q(wv)
    \qquad\text{and}\qquad p(wav) = q(wav)
  \end{equation}
hold for all choices of $v\in I, w\in J$ and $a\in\S$.
\end{itemize}
\end{theorem}

\Pf Given a basic matrix $P_{IJ}$ together with the probabilities
$p(v),p(wav)$ for all $v\in I, w\in J, a\in\S$, one can reconstruct
$p$ via a ''minimal representation'' (see, \emph{e.g.},
\cite{Ito92,Jaeger00,Schoenhuth07} for details).\qed\\

\section{Parametrizations and The Equivalence Problem}
\label{sec.equivalence}

\subsection{Hidden Markov Processes}
\label{ssec.hmms}

%A \emph{hidden Markov model} is a tuple $\cM=(S, X, \pi, M)$ where
%\begin{enumerate}
%\item $S=\{s_1,\ldots,s_n\}$ is a finite set of \emph{(hidden) states}
%  and $X:S\to \Sigma$ a map.
%\item $\pi$ is a probability distribution on $S$ and $M=[m_{ij}]\in
%  \R^{S\times S}$ a non-negative matrix with unit column sums
%  $\sum_{i=1}^n m_{ij} =1$ (\emph{i.e.} the column vectors of $M$ are
%  probability distributions on $S$).
%\end{enumerate}
%
%The associated \emph{hidden Markov process} $(X_t)$ moves to a state
%$s\in S$ with probability $\pi^{(0)}_s$ and emits the symbol $X_1 =
%X(s)$. Then it moves from $s$ to a state $s'$ with probability
%$m_{ss'}$ and emits the symbol $X_2=X(s')$ and so on. We call $\cM$ a
%\emph{parametrization} of $(X_t)$.

A \emph{hidden Markov process (HMP)} is parametrized by a tuple $\cM=(S, E,
\pi, M)$ where
\begin{enumerate}
\item $S=\{s_1,\ldots,s_n\}$ is a finite set of \emph{``hidden'' states}
\item $E=[e_{sa}]\in\R^{S\times \S}$ is a non-negative {\em emission
  probability matrix} with unit row sums $\sum_{a\in\S}E_{sa}=1$,
  (\emph{i.e.} the row vectors of $E$ are probability distributions on
  $\S$)
\item $\pi$ is an \emph{initial probability distribution} on $S$ and 
\item $M=[m_{ij}]\in\R^{S\times S}$ is a non-negative \emph{transition
  probability matrix} with unit row sums $\sum_{i=1}^nm_{ij} =1$
  (\emph{i.e.} the row vectors of $M$ are probability distributions on
  $S$
%\footnote{For technical convenience, we opted to define $M$
%    with columns, instead of rows, as probability distributions}).
\end{enumerate}

The associated process $(X_t)$ initially moves to a state $s\in S$
with probability $\pi_s$ and emits the symbol $X_1=a$ with probability
$E_{sa}$. Then it moves from $s$ to a state $s'$ with probability
$m_{ss'}$ and emits the symbol $X_2=a'$ with probability $e_{s'a'}$
and so on. In the following, we also refer to a parametrization
$\cM=(S,E,\pi,M)$ as a \emph{hidden Markov model (HMM)}.

%\begin{remark}
%Among practitioners, parametrizations of the type $\cM=(S, E, \pi, M)$
%where the map $X:S\to \S$ is replaced by a {\em emission probability
%  matrix} $E\in\R^{S\times\S}$ might be more prevalent. In that case
%each entry $E_{sa},s\in S,a\in\S$ reflects the probability to emit
%symbol $a$ from hidden state $s$ and the process generates symbols by
%traveling the hidden states as usual and generating symbols $a$ with
%probabilities $E_{sa}$ from the hidden states $s$.\par Note that both
%classes of parametrizations give rise to the same class of stochastic
%processes \cite{Gilbert59}. Here we opted to treat parametrizations of
%the first type which is in line with earlier work on related topics
%\cite{Gilbert59,Heller65,Ito92,FaiSchoe07}. Note furthermore that this
%does not affect any runtime considerations. See remark
%\ref{rem:hmmruntime} for explanations.
%\end{remark}

\subsection{Quantum Random Walks}
\label{ssec.qrws}

A \emph{quantum random walk (QRW)} is parametrized by a tuple
$\cQ=(G,U,\psi_0)$ where
\begin{enumerate}
\item $G=(\S,E)$ is a directed, $K$-regular graph over the alphabet $\S$
\item $U:\C^k\to\C^k$ is a unitary \emph{evolution} operator where
  $k:=|E|=K\cdot|\S|$ and
\item $\psi_0\in\C^k$ is a wave function, that is $||\psi_0||=1$
  ($||.||$ the Euclidean norm).
\end{enumerate}
Edges are labeled by tuples $(a,x), a\in\S, x\in X$ where $X$ is
a finite set with $|X|=k$. Correspondingly, $\C^k$ is considered to be
spanned by the orthonormal basis 
\begin{equation*}
\langle\; \ve_{(a,x)}\,|\,(a,x)\in E \;\rangle. 
\end{equation*}
According to \cite{Aharonov01} some more specific conditions must hold
which do not affect our considerations here.\\

The \emph{quantum random walk} $(X_t)$ arising from a parametrization
$\cQ=(G,U,\psi_0)$ proceeds by first applying the unitary operator $U$
to $\psi_0$ and subsequently, with probability $\sum_{x\in
  X}|(U\psi_0)_{(a,x)}|^2$, ``collapsing'' (i.e.~projecting and
renormalizing, which models a quantum mechanical measurement)
$U\psi_0$ to the subspace spanned by the vectors $\ve_{(a,x),x\in X}$
to generate the first symbol $X_1=a$. Collapsing $U\psi_0$ results in
a new wave function $\psi_1$. Applying $U$ to $\psi_1$ and collapsing
it, with probability $\sum_{x\in X}|(U\psi_1)_{(a',x)}|^2$, to the
subspace spanned by $\ve_{(a',x),x\in X}$ generates the next symbol
$X_2=a'$. Iterative application of $U$ and subsequent collapsing
generates further symbols.

\subsection{The Equivalence Problem}
\label{ssec:equivalence}

The \emph{equivalence problem} can be framed as follows:\par
\bigskip
\hrule
\begin{center}{\bf Equivalence Problem (IP)}\end{center} 
\vspace{-1ex} Given two hidden Markov models $\cM_X,\cM_Y$ or two
quantum random walk parametrizations $\cQ_X,\cQ_Y$, decide whether the
associated processes $(X_t)$ and $(Y_t)$ are equivalent.\\[-.5ex]

\hrule
\bigskip

%The \emph{equivalence problem} for Markov processes is:
%
%\medskip
%\begin{itemize}
%\item[$\bullet$] Given two hidden Markov models $\cM$ and $\cM'$,
%  decide whether the associated Markov processes $(X_t)$ and $X_t')$
%  have the same probability distributions.
%\end{itemize}

\medskip
The equivalence problem can, of course, be solved in principle, in the
spirit of Theorem \ref{t.key-theorem}. In order to efficiently solve
it in practice, it suffices to be able to efficiently compute the
following quantities:

\begin{itemize}
\item[(1)] A basis $(I,J)$ for the finitary processes $(X_t),(Y_t)$
  from their parametrizations $\cM_X,\cM_Y$.
\item[(2)] The corresponding probabilities $p(v), p(wv), p(wav)$ for
  all choices of $v\in I,w\in J,a\in\S$.
\end{itemize}

\section{Computing Probabilities}
\label{sec.computations}

We would like to point out that in the following we assume that all
inputs consist of rational numbers and that each arithmetic operation
can be done in constant time. This agrees with the usual conventions
when treating related probabilistic concepts in terms of algorithmic
complexity \cite{Rabiner89,Tzeng92,Dupont05}.

\subsection{Hidden Markov Processes}
\label{ssec.hmm-computations}

Let now $(X_t)$ be a hidden Markov process with parametrization
$\cM=(S,E,\pi,M)$.  Observe first that the \emph{transition matrix}
$M$ decomposes as $M=\sum_{a\in \Sigma} T_a$ into matrices $T_a$ with
coefficients
\begin{equation}
    (T_a)_{ij} := e_{s_ia}\cdot m_{ij}
    %\left\{\begin{array}{cl} m_{ij}\cdot E_{s_ja} 
    %&\mbox{if $X(j)=a$}\\
    %0 &\mbox{otherwise.}\end{array}\right.
\end{equation}
which reflect the probabilities to emit symbol $a$ from state $s_i$ and
subsequently to move on to state $s_j$.
%\begin{sloppypar}
Standard technical computations (\emph{e.g.} \cite{FaiSchoe07})
reveal that that for any word $a_1\ldots a_t\in\Sigma^t$:%\end{sloppypar}
\begin{equation}
\begin{split}
p(a_1a_2\ldots a_t) &= p(a_1a_2\ldots a_{t-1})p(a_t| a_1a_2\ldots a_{t-1})\\
&= \cdots \\
&= \pi^TT_{a_1}\ldots T_{a_{t-1}}T_{a_t}\mathbf{1},
\end{split}
\end{equation}
where $\mathbf{1}=(1,...,1)^T\in\R^S$ is the vector of all ones.\\  

For further reference, we use the notations
\begin{eqnarray}
T_v &:=& T_{v_1}T_{v_2}\ldots T_{v_{t-1}}T_{v_t}\in\R^{n\times n} 
\end{eqnarray}
for any $v=v_1\ldots v_t\in \Sigma^*$ as well as
\begin{equation}
\ora{\vp}(v) := \pi^TT_v\in\R^{1\times n}\qquad\text{and}\qquad
\ola{\vp}(v) := T_v\mathbf{1}\in\R^{n\times 1}.
\end{equation}

\medskip
\begin{remark}
\label{rem.forwardbackward}
Note that computation of vectors $\ora{\vp}(v)$ and $\ola{\vp}(v)$
%\begin{equation}
%\pi^TT_{a_1}...T_{a_s} \quad\text{and}\quad T_{a_{s+1}}...T_{a_{t}}\mathbf{1}
%\end{equation}
is an alternative way to describe the well-known Forward and Backward
algorithm (\emph{e.g.} \cite{Ephraim02}) since the entries of these
two vectors can be identified with the Forward and Backward variables
\begin{gather}
\Pr(S_{s+1} = s_i\,|\, X_1=a_1,...,X_s=a_s)\\
\text{and}\\
\Pr(S_{s+1}=s_i\,|\,X_{s+1}=a_{s+1},...,X_{s+t}=a_{s+t})
\end{gather}
where $(S_t)$ is the (non-observable) Markov process over the hidden
states $S=(s_1,...,s_n)$.
\end{remark}

\subsection{Quantum Random Walks}
\label{ssec.qrw-computations}

The following considerations can be straightforwardly derived from
standard quantum mechanical arguments, see \cite{Nielsen} for 
a reference.

\subsubsection{The State Space $\cS^n$}

We write $Q^*$ for the adjoint of an arbitrarily sized matrix
$Q\in\C^{m\times n}$, (that is $Q^*_{ji}= a - ib$ if $Q_{ji}= a + ib$
where usage of $i$ as both a running index and a complex number should
not lead to confusion).  Let 
\begin{equation}
n := k^2 = |E|^2.
\end{equation}
We will consider the set of self-adjoint matrices
\begin{equation}
\cS^n:=\{Q\in\C^{k^2}\,|\,Q=Q^*\}
\end{equation}
in the following, which is usually referred to as {\em state space} in
quantum mechanics. As usual, $\cS^n$ can be viewed as an
$n=k^2$-dimensional real-valued vector space. To illustrate this let
\begin{eqnarray}
\ve_m & := & (0,...,0,\underset{m}{1},0,...,0)^T\in\C^K,m=1,...,k 
\quad\text{and}\quad\\
\vf_m & := & (0,...,0,\underset{m}{i},0,...,0)^T\in\C^K,m=1,...,k.
\end{eqnarray}
The self-adjoint matrices
\begin{equation}
\label{eq.statespacebasis}
\vE_{m_1m_2} := (\ve_{m_1}\ve_{m_2}^*+\ve_{m_2}\ve_{m_1}^*)\quad\text{and}\quad
\vF_{m_1m_2} := (\vf_{m_1}\vf_{m_2}^*+\vf_{m_2}\vf_{m_1}^*)
\end{equation}
for all choices of $1\le m_1,m_2\le k$ and $m_1\ne m_2$ for
$\vF_{m_1m_2}$ (since entries on the diagonal of self-adjoint matrices
are real-valued) then form a canonical basis of $\cS^n$ (note that
$\vE_{m_1m_2}=\vE_{m_2m_1},\vF_{m_1m_2}=\vF_{m_2m_1}$).

%As each $Q\in\mathcal{S}^n$ can be written as
%\begin{equation}
%Q=A+iB,\; A,B\in\R^{n²}
%\end{equation}
%where $A$ is symmetric and $B$ is skew-symmetric, a basis for $\cS^n$
%can be composed by joining a basis of the symmetric real-valued
%matrices (such a basis has $n(n+1)/2$ elements) and a basis of the
%skew-symmetric matrices (this has cardinality $n(n-1)/2$) which
%reveals that $\cS^n$ can be viewed as a real-valued vector space of
%dimension $n^2$.\\
%
%\begin{remark}
%\label{rem.basis}
%Let ($i\in\C,i^2=1$)
%\begin{eqnarray}
%e_m & := & (0,...,0,\underset{m}{1},0,...,0)^T\in\C^n,m=1,...,n \quad\text{and}\\
%f_m & := & (0,...,0,\underset{m}{i},0,...,0)^T\in\C^n,m=1,...,n.
%\end{eqnarray}
%A canonical choice of a basis for $\cS^n$ then is
%$(e_{m_1}e_{m_2}^*+e_{m_2}e_{m_1}^*),1\le m_1,m_2\le n,
%(f_{m_1}f_{m_2}^*+f_{m_2}f_{m_1}^*),1\le m_1,m_2\le n, m_1\ne m_2$.
%Therefore, whenever we refer to a basis $(Q_k),k=1,...,n^2$ of $\cS^n$
%in the following we assume that $Q_k$ is of the form
%\begin{equation}
%Q_k = (e_{m_1}e_{m_2}^*+e_{m_2}e_{m_1}^*)\quad\text{or}\quad
%Q_k = (f_{m_1}f_{m_2}^*+f_{m_2}f_{m_1}^*).
%\end{equation}
%\end{remark}

\subsubsection{Linear Operations on $\cS^n$}

For a quantum random walk parametrization $\cQ=(G=(\S,E),U,\psi_0)$ 
we introduce the projection operators ($k:=|E|$)
\begin{equation}
P_a:\C^k\longrightarrow\C^k,\;\psi\mapsto\sum_{(a,x),x\in X}\psi_{(a,x)}\ve_{(a,x)}
\end{equation}
for all $a\in\S$ which reflects projection of $\psi$ onto the subspace
spanned by the $\ve_{(a,x)},x\in X$. We find that
\begin{equation}
\label{eq.Tu}
T_a:\cS^n\longrightarrow\cS^n,\;Q\mapsto (P_aU)Q(P_aU)^*
\end{equation}
is an $\R$-linear operator acting on the state space $\cS^n$.
In analogy to the
theory of hidden Markov models, where here the order on the
letters has been reversed, we further define
\begin{equation}
T_v := T_{v_t}T_{v_{t-1}}\ldots T_{v_2}T_{v_1}\in\R^{n\times n} 
\end{equation}
for any $v=v_1\ldots v_t\in \Sigma^*$.\\

Let now $Q_{\psi}:=\psi\psi^*\in\C^{k\times k}$ be the self-adjoint
matrix being associated with a wave function $\psi\in\C^k$. We recall
that, by definition of the quantum random walk $p$ with
parametrization $\cQ$, probabilities $p(v=v_1...v_t)$ are computed as
\begin{equation}
%\label{eq.qrwprob}
p(v=v_1...v_t)
= ||(P_{v_t}U)(P_{v_{t-1}})\hdots(P_{v_1}U)\psi_0||^2
%= \tr (P_{v_t}U)\hdots(P_{v_1}U)Q_{\psi_0}(P_{v_1}U)^*\hdots(P_{v_t}U)^*.
\end{equation}
which can be rephrased as ($Q_{\psi_0}:=\psi_0\psi_0^*$ and tr is
the linear trace functional, that is the sum of the diagonal entries)
\begin{equation}
\label{eq.qrwprob}
p(v=v_1...v_t)
%= ||(P_{v_t}U)(P_{v_{t-1}})\hdots(P_{v_1}U)\psi_0||
%= \tr (P_{v_t}U)\hdots(P_{v_1}U)Q_{\psi_0}(P_{v_1}U)^*\hdots(P_{v_t}U)^*
= \tr T_{v_t}...T_{v_1}Q_{\psi_0}
\end{equation}
which yields that probabilities $p(v)$ can be computed by iterative
application of multiplying $n\times n$-matrices with $n$-dimensional
vectors where we recall that $Q_{\psi_0}$ can be taken as an element
of the $n$-dimensional vector space $\cS^n$. Note that $T_v$ acts on
$Q_{\psi_0}$ in the sense of $\cS^n$ whereas the trace functional
treats $T_vQ_{\psi_0}$ as a matrix.

\subsubsection{Forward and Backward Algorithm}

Note that application of the trace functional can be rephrased
as 
\begin{equation}
\tr Q = E\cdot Q\in\R\quad\text{where}\quad E:=\sum_{i=1}^k\ve_i\ve_i^*
\end{equation}
and, on the right hand side, both $E$ and $Q$ are taken as
elements of $\cS^n$, i.e.~as $n$-dimensional vectors.  Using this, we define 
\begin{equation}
\ora{\vp}(v) := T_vQ_{\psi_0}\in\cS^n\subset\C^{n^2}\qquad\text{and}\qquad
\ola{\vp}(v) := ET_v\in\C^{n^2}.
\end{equation}
Computation of $\ora{\vp}(v)$ and $\ola{\vp}(v)$ can be taken as
performing a quantum random walk version of the Forward and the
Backward algorithm. Correspondingly, entries of $\ora{\vp}(v)$ and
$\ola{\vp}(v)$ reflect Forward and Backward variables.

\subsection{Runtimes}
\label{ssec.runtimes}

Since the multiplication of an $(n\times n)$-matrix with a vector can
be done in $O(n^2)$ arithmetic operations, the previous considerations
let us conclude:

\begin{lemma}
\label{l.probability-computation}  
Given $\cM$ or $\cQ$ let $n$ be the number of hidden states $|S|$
resp.~the dimension of the state space $\cS^n$ associated with $\cQ$
and $p$ be the probability function of $\cM$ or $\cQ$.
\begin{enumerate} 
\item 
For any $v\in \Sigma^*$
\begin{equation}
\ora{\vp}(v),\ola{\vp}(v)\quad\text{and}\quad p(v)
\end{equation}
can be computed in $O(|v|n^2)$ arithmetic operations.
\item
Upon computation of $\ora{\vp}(w)$ computation of all
\begin{equation}
\label{eq.ora}
p(wa)\quad\text{and}\quad\ora{\vp}(wa)
%,\ola{\vp}(va)\quad\text{and}\quad p(av),\ora{\vp}(av),\ola{\vp}(av) 
\end{equation}
requires $O(|\S|n^2)$ arithmetic operations.
\item
Upon computation of 
%$\ora{\vp}(v)$ and
$\ola{\vp}(v)$ computation of all 
\begin{equation}
\label{eq.ola}
%p(va),\ora{\vp}(va),\ola{\vp}(va)
%\quad\text{and}\quad 
p(av)\quad\text{and}\quad\ola{\vp}(av) 
\end{equation}
requires $O(|\S|n^2)$ arithmetic operations.
\item
Upon computation of $\ora{\vp}(w)$ and
$\ola{\vp}(v)$ computation of all 
\begin{equation}
\label{eq.wav}
p(wav)
%,\ora{\vp}(va),\ola{\vp}(va)\quad\text{and}\quad
%p(av),\ora{\vp}(av),\ola{\vp}(av)
\end{equation}
requires $O(|\S|n^2)$ arithmetic operations.
\qed
\end{enumerate}
\end{lemma}

For hidden Markov models $\cM$ this actually reflects
well-known results on computation of Forward/Backward variables.

\section{Equivalence Tests}
\label{sec.equivalencetests}

In this section, we describe how to efficiently test two hidden Markov
processes or quantum random walks $(X_t)$ and $(Y_t)$ for equivalence.
We recall that a generic strategy has been established by theorem
\ref{t.key-theorem}. Our solution proceeds according to this strategy.

\subsection{Computation of a Basis} 
\label{ssec.basis-computation}

We will now show how to compute a basis $(IJ)$ in runtime $O(|\S|n^4)$
for a hidden Markov process resp.~a quantum random walk
$p$. Therefore, assume for now that $g_1,\ldots,g_n:\Sigma^*\to \R$
are probability functions the probabilities of which can be computed
in the style of hidden Markov processes resp.~quantum random walks and
which generate the column space of $p$, \emph{i.e.},
\begin{equation}
     \cC(p) \;\subset\; \spann \{g_1,\ldots,g_n\}.
\end{equation}
Given $g_1,\ldots,g_n$, computation of a basis $(IJ)$ proceeds in
three steps the first two of which are analagous and the third of
which is a simple procedure.
\begin{enumerate}
\item Compute a row generator $I$.
\item Compute a column basis $J$.
\item Reduce $I$ to a row basis.
\end{enumerate}
While steps $1$ and $2$ both require runtime $O(|\S|n^4)$, step $3$
requires $O(n^4)$ which overall evaluates as $O(|\S|n^4)$ runtime
required for computation of a basis.\\

We discuss the steps in the following paragraphs. In a subsequent
subsection, we show how to obtain suitable $g_1,...,g_n$ for both
hidden Markov models and quantum random walks.

\subsubsection{Step 1: Computation of a row generator $I$}

Consider the following algorithm.

\medskip
\begin{Alg}
\label{a.I}
$\;$
\end{Alg}
\vspace{-1ex}

\begin{algorithmic}[1]
\STATE Define $\vg(v)=(g_1(v),...,g_n(v))\in\R^{n}$.
\STATE $I \la \{\Box\}, B_{row}\la \{\vec{g}(\Box)\}, C_{row}\la\Sigma$.
\WHILE{$C_{row}\ne\emptyset$}
\STATE Choose $v\in C_{row}$.
\IF{$\vg(v)$ is linearly independent of $B_{row}$}
\STATE $I\la I\cup\{v\}, B_{row}\la B_{row}\cup\{\vg(v)\}$\\
       $C_{row}\la C_{row}\cup\{av\,|\,a\in\Sigma\}$
\ENDIF
\ENDWHILE

\STATE {\bf output} $I$.
\end{algorithmic}
%\end{Alg}

\medskip
\begin{proposition}
\label{p.algorithm1} 

Let $I\subseteq \Sigma^*$ be the output of
Algorithm~\ref{a.I}. Then one has
\begin{equation}
\label{eq.propstate}
\cR(p) = \spann \{p_v\mid v\in I\} 
\quad\text{and}\quad\dim(X_t)\le |I|
%\begin{cases} = |I|, & \cC(p) = \spann\{g_1,\ldots,g_n\}\\
%              < |I|, & \cC(p) \subsetneq \spann \{g_1,\ldots,g_n\}.
%\end{cases}
\end{equation}
where 
\begin{equation}
\label{eq.dimspann}
\cC(p) = \spann \{g_1,\ldots,g_n\} \quad\Rightarrow\quad \dim(X_t)=|I| .
\end{equation}
Furthermore, 
\begin{enumerate}
\item[(i)] The algorithm terminates after at most $|\Sigma|\cdot n$
  iterations.
\item[(ii)] Each iteration requires $O(n^3)$ arithmetic operations where
  at most $n$ iterations need additional $O(|\S|n^3)$ operations.
\end{enumerate}
\end{proposition}

\Pf Ad $(i)$: Because the $n$-dimensional vectors in $B_{row}$ are
independent $|B_{row}|\le n$ and $|I|\leq n$ follow immediately.
%Since the lengths of the words in $C_{row}$ increase at most by
%$1$ in each step where a new row $\vg(v)$ is added, $|v|\leq |I|\leq
%n$ follows for all $v\in I$. 
Since at most $\Sigma$ words are added to $C_{row}$ upon discovery of
an $n$-dimensional vector which is linearly independent of those in
$B_{row}$, we have $|C_{row}|\le|\Sigma|\cdot n$ and hence at most
$|\Sigma|\cdot n$ iterations.

Ad $(ii)$: In each iteration, we perform a test for linear
independency of at most $n$ vectors of dimension $n$ which requires at
most $O(n^3)$ arithmetic operations \cite{Faddeev63}. In the at most
$n$ cases where $\vg(v)$ is linearly independent of $B_{row}$, we
proceed by computing
\begin{equation}
(g_1(av),...,g_n(av))\quad\text{and}\quad(\ola{\vg_1}(av),...,\ola{\vg_n}(av))
\end{equation}
for all $a\in\S$ where
$(\ola{\vg_1}(v),...,\ola{\vg_n}(v))$ are
available from an iteration before (note that
$g_i(\Box)=1,\ola{\vg_i}(\Box)=(1,...,1)$ in the first iteration).
Due to lemma \ref{l.probability-computation}, (\ref{eq.ola}),
this requires $O(|\S|\cdot n^3)$ operations.\\

To prove (\ref{eq.propstate}), let $w_0\in \Sigma^*$ be arbitrary and
suppose
\begin{equation}
p_{w_0}\notin\spann\{p_v\mid v\in I\}.
\end{equation}
Since $\cC(p)\;\subset\;\spann\{g_1,...,g_n\}$, plugging $w=\Box$ into
lemma~\ref{l.hilfslemma} below implies
\begin{equation}
  \vg(w_0) \notin \spann\{\vg(v)\mid v\in I\}.
\end{equation}
We will derive a contradiction. Indeed, the algorithm can only miss
$w_0$ if $w_0$ had never been collected into $C_{row}$ in step 6.
This happens only in case that there is a $v_0\in\Sigma^*$ such that
\begin{equation}
w_0 = wv_0 
\end{equation}
holds for some $w\in\Sigma^*$ and $\vg(v_0)$ had been found
to be linearly dependent of $[\vg(v)]_{v\in I}$. Lemma \ref{l.hilfslemma}
below then states that in such a case $p_{w_0}\in\spann\{p_{wv}\mid v\in I\}$
holds and it remains to show that for each $w\in\S^*$ and $v\in I$
\begin{equation}
\label{eq.remainswv}
p_{wv}\in\spann\{p_v\mid v\in I\}. 
\end{equation}
This follows by induction on the length $|w|$ of $w$ from the
following arguments.  For each $w\in\S^*$ we define a linear operator
$\sigma_w$ on $\cR(p)$ through
\begin{equation}
\sigma_wp_v = p_{wv}.
\end{equation}
By design of the update rule for $C_{row}$ in step 6 of algorithm \ref{a.I} we
immediately see that
\begin{equation}
\vg(av)\in\spann\{\vg(v)\mid v\in I\}
\end{equation}
for all $a\in\S$, hence by plugging $v_0=av$ and $w=\Box$ into
lemma~\ref{l.hilfslemma}, we obtain $p_{av}\in\spann\{p_v\mid v\in
I\}$ that is
\begin{equation}
\sigma_a(\spann\{p_v\mid v\in I\})\;\subset\;\spann\{p_v\mid v\in I\}
\end{equation}
for all $a\in\S$. Inductively, by observing that
$\sigma_{w=w_1...w_t}=\sigma_{w_1}\circ...\circ\sigma_{w_t}$,
\begin{equation}
\sigma_w(\spann\{p_v\mid v\in I\}\;\subset\;\spann\{p_v\mid v\in I\}
\end{equation}
and thereby (\ref{eq.remainswv}).\\

To see (\ref{eq.dimspann}) let $\dim(X_t)<|I|$. Since $|I|=|B_{row}|$
we obtain that
\begin{equation}
\dim\cC(p)=\dim(X_t)<|B_{row}|\le\dim\spann\{g_1,\ldots,g_n\}
\end{equation}
hence
$\cC(p)\subsetneq\spann\{g_1,\ldots,g_n\}$. \qed\\

\begin{lemma}
\label{l.hilfslemma}
Let $g_1,\ldots,g_n:\Sigma^*\to \R$ be such such that $\cC(p) \subseteq
\spann\{g_1,\ldots,g_n\}$ and let $v_0,v_1,...,v_m\in \Sigma^*$ be
such that
\begin{equation}
\label{eq.gv}
(g_1(v_0),...,g_n(v_0))
\in\spann\{(g_1(v_j),...,g_n(v_j))\,|\,j=1,...,m\}\subseteq\R^n.
\end{equation}
Then one has for every $w\in \Sigma^*$:
\begin{equation}
\label{eq.colv0vw}
p_{wv_0}\in\spann\{p_{wv_j}\,|\,j=1,...,m\}\subseteq \R^n.
\end{equation}

The analogous statement holds for the row space $\cR(p)$.
\end{lemma}

\Pf  By our hypothesis, there are scalars $\beta_1,...,\beta_m\in \R$ such that
\begin{equation}
\label{eq.betaj}
(g_1(v_0),...,g_n(v_0)) = \sum_{j=1}^m\beta_j(g_1(v_j),...,g_n(v_j)).
\end{equation}
Let $u\in\Sigma^*$ be arbitrary. Again by our hypothesis, there are scalars
$\alpha_i,i=1,...,n\in \R$ such that
\begin{equation}
\label{eq.alphai}
p_u = \sum_{i=1}^n\alpha_ig_i.
\end{equation}
We now compute
\begin{eqnarray*}
p_u(v_0)  &&\stackrel{(\ref{eq.alphai})}{=} \sum_{i=1}^n\alpha_ig_i(v_0)\\
&&\stackrel{(\ref{eq.betaj})}{=} \sum_{i=1}^n\alpha_i\sum_{j=1}^m\beta_jg_i(v_j)
= \sum_{j=1}^m\beta_j\sum_{i=1}^n\alpha_ig_i(v_j)\\
&&\stackrel{(\ref{eq.alphai})}{=} \sum_{j=1}^m\beta_jg_u(v_j) = \sum_{j=1}^m\beta_jp(uv_j)
= \sum_{j=1}^m\beta_j p_{v_j}(u).
\end{eqnarray*}
Since the $\beta_j$ had been determined independently of $u$, we thus conclude
\begin{equation}
\label{eq.fv0fvj}
p_{v_0} = \sum_{j=1}^m\beta_j p_{v_j}.
\end{equation}
Let $\sigma_w$ be the linear operator on $\cR(p)$  with the property
\begin{equation}
      \sigma_w p_v = p_{wv}.
\end{equation}

Application of $\sigma_w$ to (\ref{eq.fv0fvj}) then shows
\begin{equation}
p_{v_0w} = \sigma^w(p_{v_0}) = \sum_{j=1}^m\beta_j \sigma_w( p_{v_j}) =
\sum_{j=1}^m\beta_j p_{v_jw},
\end{equation}
which implies (\ref{eq.colv0vw}).\qed\\

\subsubsection{Step 2: Computation of a column basis $J$}

Having obtained the row generator $I\subseteq \Sigma^*$ in the step
before, that is $\dim (X_t)\le |I|$ and
\begin{equation}
     \cR(p) = \spann \{p_v\mid v\in I\},
\end{equation}
we can now use these functions $p_v$ as an input for an algorithm
which is analogous to that for computing the row generator $I$.

\medskip
\begin{Alg}
\label{a.J}
$\;$
\end{Alg}
\vspace{-1ex}

\begin{algorithmic}[1]
\STATE Define $\vec{q}(w):=(p_v(w)=p(wv),v\in I)\in\R^{|I|}$.
\STATE $J \la \{\Box\}, B_{col}\la \{\vec{q}_w(\Box)\}, C_{col}\la\S$
\WHILE{$C_{row}\ne\emptyset$}
\STATE Choose $w\in C_{col}$.
\IF{$\vec{q}(w)$ is linearly independent of $B_{col}$}
\STATE $A_{col}\la A_{col}\cup\{w\}, B_{col}\la B_{col}\cup\{q(w)\}$\\ 
       $C_{col}\la C_{col}\cup\{wa\,|\,a\in\S\}$
\ENDIF
\ENDWHILE
%\IF{$|A_{row}|>|A_{col}|$}
%\STATE Eliminate $|A_{col}|-|A_{row}|$ many strings from $A_{row}$ such
%       that $[p(wv)]_{v\in A_{row},w\in A_{col}}$ is regular for the 
%       remaining $v\in A_{row}$.
%\ENDIF
\STATE {\bf output} $J$%$V:=[p(wv)]_{v\in A_{row},w\in A_{col}}$.
\end{algorithmic}

\medskip
While this routine is, in essence, analogous to algorithm \ref{a.I},
there is one difference to be observed: Here $C_{col}$ gets augmented
by joining $wa$ whereas $C_{row}$, in algorithm \ref{a.I}, was
augmented by joining $av$. This asymmetry is due to that one obtains
an equivalently asymmetric statement in lemma \ref{l.hilfslemma} when
rephrasing it for $\cR(p)$ instead of $\cC(p)$. As a consequence,
application of (\ref{eq.ora}) instead of (\ref{eq.ola}) in lemma
\ref{l.probability-computation} is needed.\\

We obtain that
\begin{equation}
        P_{IJ} =[p(wv)]_{v\in I,w\in J}
\end{equation}
is a generator for $(X_t)$. Since
$\cR(p)=\spann\{p_v\mid v\in I\}$, by applying (\ref{eq.dimspann}),
we see that
\begin{equation}
|J| = \dim(X_t).
\end{equation}
Hence $J$ is a genuine column basis. We recall that this
was not necessarily the case for $I$ which can happen to occur
in the case $\cC(p) \subsetneq \spann\{g_1,...,g_n\}$.\\

All $p(wav),v\in I,w\in J,a\in \S$ can be obtained in runtime
$O(|\S|\cdot n^4)$ through application of (\ref{eq.wav}) in lemma
\ref{l.probability-computation} making use of the
$\ora{\vp}(w),\ola{\vp}(v)$ which were computed when executing the
algorithms \ref{a.I}, \ref{a.J}.

We conclude: all necessary quantities can be obtained through
$O(|\S|\cdot n^4)$ arithmetic operations.

\subsubsection{Step 3: Making $I$ a basis}

This step is simple: one removes $v$ from $I$ where $p(wv),w\in J$ is
linearly dependent in $P_{IJ}$. This reduces the possibly too large
set $I$ to a row basis and finally yields a basis $(IJ)$ for
$(X_t)$. This requires at most $n$ linear independence tests of
$n$-dimensional vectors hence $O(n^4)$ runtime \cite{Faddeev63}.

\subsection{Generating sets}\label{sec:Generating-sets} 

Let us call a set $\{g_1,\ldots,g_n\}$ of functions $g_i$ as in the
previous section a set of \emph{generators} for the column space
$\cC(p)$ of the hidden Markov process resp.~quantum random walk
$(X_t)$.\\

We can get sets of generators as follows for which probabilities
$g_i(v)$ can be computed in the style of hidden Markov processes
resp.~quantum random walks as follows.

\subsubsection{Hidden Markov Processes}

Given a hidden Markov model $\cM=(S,E,\pi,M)$, consider the hidden
Markov models $\cM_i =(S,X,\ve_i,M)$, where $\ve_i$ is the $i$th unit
vector in $\R^S$. One now takes
\begin{equation}
    g_i(v) = \ve_i^TT_v\mathbf{1} \quad(i=1,\ldots,n).
\end{equation}

\subsubsection{Quantum Random Walks}

For a quantum random walk, as parametrized through a self-adjoint
matrix $Q_{\psi_0}$ and linear operators $T_v,v\in\S^*$ (acting
on the state space see subsection \ref{ssec.qrw-computations}), we
see that
\begin{equation}
  g_i(v) = \tr T_vQ_i \quad(i=1,\ldots,n)
\end{equation}
where the $Q_i$ comprise all of the state space basis members
$\vE_{m_1m_2},\vF_{m_1m_2}$ (see (\ref{eq.statespacebasis})).

\subsection{Summary}
\label{ssec.summary}

Theorem \ref{t.key-theorem} yields the following procedure as an
efficient test for equivalence of processes $(X_t)$ and $(Y_t)$, :
\begin{enumerate}
\item Compute a basis for both $(X_t)$ and $(Y_t)$.
\item If $\dim(X_t)\ne\dim(Y_t)$ return {\bf not equivalent}.
\item If $\dim(X_t)=\dim(Y_t)$, perform equality tests from
  (\ref{eq.keyequalities}).
\item Output {\bf equivalent} if all of them apply and {\bf not
  equivalent} if not.
\end{enumerate}

According to the above considerations, Step $1$ can be performed in
$O(|\S|n^4)$ runtime where
\begin{equation}
\label{eq.n}
n=\max\{n_X,n_Y\}
\end{equation}
and $n_X,n_Y$, in case of hidden Markov processes $(X_t),(Y_t)$, are
the numbers of hidden states and in case of quantum random walks
$(X_t),(Y_t)$ are the dimensions of the associated state spaces. For
step $2$ we recall that all strings participating in the bases, as
computed through algorithms \ref{a.I},\ref{a.J}, emerge as extensions
of basis strings obtained in an earlier iterations.  Application of
(\ref{eq.ora},\ref{eq.ola},\ref{eq.wav}) from lemma
\ref{l.probability-computation} then yields that all of the equality
tests can be equally performed in $O(|\S|n^4)$ arithmetic operations.\\

These insights can be condensed into the following main theorem where
$n$ as in (\ref{eq.n}).

\begin{theorem} 
The equivalence problem can be algorithmically solved for both hidden
Markov processes and quantum random walks in $O(|\S|n^4)$ arithmetic
operations.\qed
\end{theorem}

\subsubsection{Probabilistic Automata}

Our solution can be straightforwardly adapted to determine equivalence
of probabilistic automata which we will describe in the following. It
can therefore be viewed as more general than the main result obtained
in \cite{Tzeng92}. The main difference one has to keep in mind is that
probabilistic automata induce probability distributions on the
(countable) set of strings $\S^*$ whereas HMMs give rise to stochastic
processes, in other words to probability distributions on the
(uncountably infinite) set of sequences $\S^{\N}$. in case of
probabilistic automata equivalence then translates to equality of the
associated probability distributions on $\S^*$. The following
notations are adopted from \cite{Tzeng92}.

\begin{corollary}
\label{c.probautomata}
Let $\cA_1=(S_1,\S,M_1,\pi_1,F_1),\cA_2=(S_2,\S,M_2,\pi_2,F_2)$ be two
probabilistic automata where $N_1=|S_1|,N_2=|S_2|$. Then equivalence
of $\cA_1,\cA_2$ can be determined in $O((|\S|+1)N^4)$ where
$N=\max(N_1,N_2)$.
\end{corollary}

{\bf Proof}.$\;$ By adding a special symbol $\$$ to $\S$ which is
emitted from the final states with probability $1$ the automata
$\cA_1,\cA_2$ can be transformed into probabilistic automata with no
final probabilities $\bar{\cA_1},\bar{\cA_2}$. Let
$p_{\bar{\cA_1}},p_{\bar{\cA_2}}$ be the resulting stochastic
processes. According to \cite{Dupont05}, lemmata $3-5$, proposition
$8$, probabilistic automata with no final probabilities can be
can be viewed as HMMs $\cM_1,\cM_2$ which translates to
that for each $v\in\S^*$
\begin{equation}
p_{\cA_1}(v) = p_{\cM_1}(v)\quad\text{and}\quad p_{\cA_2}(v) = p_{\cM_2}(v).
\end{equation}
Note that the transformation from $\cA_{1,2}$ to $\cM_{1,2}$ requires
only constant time.  Applying theorem \ref{t.mainresult} to
$\cM_1,\cM_2$ yields the result.\qed\\

In short, corollary scales down the runtime $O((N_1+N_2)^4)$ (the
size of the alphabet $|\S|$ is not discussed in \cite{Tzeng92})
to $O((\max(N_1,N_2))^4)$.

\subsubsection{Ergodicity Tests}

In \cite{Schoenhuth07}, a generic algorithmic strategy for testing
ergodicity of hidden Markov processes was described, where overall
efficiency hinged on computation of a basis of the tested hidden
Markov processes. The algorithms described above resolve this
issue. Hence ergodicity of hidden Markov processes can be efficiently
tested.  Similarly to the equivalence tests, the ergodicity test of
\cite{Schoenhuth07} solely requires that the process in question is
finitary. Therefore this efficient ergodicity test equally applies for
quantum random walks.

\subsection{Conclusive Remarks}
\label{sec.conclusion}

We have presented a polynomial-time algorithm by which to efficiently
test both hidden Markov processes and quantum random walks for
equivalence. Previous solutions available for hidden Markov processes
had runtime exponential in the number of hidden states. To test
equivalence for quantum random walks, that is random walk models to be
emulated on quantum computers, is relevant for the same reasons that
apply for hidden Markov processes. An algorithm for testing
equivalence for quantum random walks had not been available
before. Note that the algorithm presented here is easy to implement
and, in particular for hidden Markov processes, only requires
invocation of well-known standard routines. Future directions are to
explore how to efficiently test for \emph{similarity} of hidden Markov
processes and quantum random walks where similarity is measured in
terms of \emph{approximate equivalence}. Such tests have traditionally
been of great practical interest.

%Overall, plugging the subroutine to determine $V$
%(algorithm \ref{a.determineV}) into algorithm \ref{a.minrepr} results
%in an algorithm which is easy to implement.
%\par Beyond having
%established a solution of the EP, we would like to mention that {\em
%  ergodicity} of an HMM or QRW, according to \cite{Schoenhuth07}, is
%equivalent to that the dimension of the eigenspace of
%$M:=V^{-1}(\sum_{a\in\S}W_a)$ (see (\ref{eq.V}),(\ref{eq.Wa})) to the
%eigenvalue $1$ is one. By means of the presented algorithm, this
%can be efficiently tested. We expect that, in the style of the
%efficient test for ergodicity, efficient tests also for other criteria
%being relevant for the theory of HMMs and QRWs can be established.

%\medskip
%\begin{remark} 
%For the class of parametrizations with rational input parameters, the
%algorithm is also polynomial in $|\Sigma|$ and the number of hidden
%states in the Turing model for computational complexity.
%\end{remark}


\begin{thebibliography}{999}

\bibitem{Aharonov01}
D.~Aharonov, A.~Ambainis, J.~Kempe, U.~Vazirani:
``Quantum walks on graphs'',
\emph{Proc. of 33rd ACM STOC, New York}, pp.~50--59, 2001.

\bibitem{Ambainis05}
A.~Ambainis:
``Quantum search algorithms'',
{\em SIGACT News}, vol.~35(2), pp.~22-35, 2004.

\bibitem{Blackwell57}
D.~Blackwell and L.~Koopmans:
``On the identifiability problem for functions of finite Markov chains'',
{\em Annals of Mathematical Statistics}, vol.~28, pp.~1011--1015, 1957.

\bibitem{Nielsen}
M.A.~Nielsen and I.L.~Chuang:
``Quantum Computation and Quantum Information'',
\emph{Cambridge University Press}, Cambridge, UK, 2000.

\bibitem{Dupont05}
P.~Dupont, F.~Denis and Y.~Esposito:
``Links between probabilistic automata and hidden Markov models: probability
distributions, learning models and induction algorithms'',
{\em Pattern Recognition}, vol.~38, pp.~1349--1371, 2005.

\bibitem{Durbin}
Durbin, Eddy, Krogh:
``Biological Sequence Analysis''
Cambridge University Press, 1998 (XXX: check)

\bibitem{Ephraim02}
Y.~Ephraim and N.~Merhav:
``Hidden Markov Processes'',
{\em IEEE Transactions on Information Theory}, vol.~48, 1518-1569, 2002.


\bibitem{FaiSchoe07}
U.~Faigle and A. Schoenhuth:
''Asymptotic mean stationarity of sources with finite evolution dimension'',
{\em IEEE Transactions on Information Theory}, vol.~53, 2342-2348, 2007.

\bibitem{FaiSchoe10}
U.~Faigle and A. Schoenhuth:
''Discrete Quantum Markov Chains'',
Preprint 2010, submitted.

\bibitem{Faddeev63}
D.~K.~Faddeev and V.~N.~Faddeeva:
``Computational Methods of Linear Algebra'',
{\em Freeman}, San Francisco, 1963.

\bibitem{Gilbert59}
E.J.~Gilbert:
``On the identifiability problem for functions of finite Markov chains'',
{\em Annals of Mathematical Statistics}, vol.~30, pp.~688--697, 1959.

\bibitem{Heller65}
A.~Heller:
``On stochastic processes derived from Markov chains'',
{\em Annals of Mathematical Statistics}, vol.~36, pp.~1286--1291, 1965.

\bibitem{Ito92}
H.~Ito, S.-I.~Amari and K.~Kobayashi:
``Identifiability of hidden Markov information sources and their minimum
degrees of freedom'', {\em IEEE Trans.~Inf.~Theory}, vol.~38(2), pp.~324--333, 1992.

\bibitem{Jaeger00}
H.~J\"ager:
``Observable operator models for discrete stochastic time series'',
{\em Neural Computation}, vol.~12(6), pp.~1371--1398, 2000.

\bibitem{Kempe05}
J.~Kempe:
``Quantum random walks hit exponentially faster'',
{\em Probability Theory and Related Fields}, vol.~133(2), pp.~215--235, 2005.

\bibitem{Marquezino08}
F.L.~Marquezino, R.~Portugal, G.~Abal and R.~Donangelo:
``Mixing times in quantum walks on the hypercube'',
\emph{Phys.~Rev.~A}, 77, 042312, 2008.

\bibitem{Merhav02}
Y.~Ephraim, N.~Merhav,
"Hidden Markov processes",
\emph{IEEE Trans. on Information Theory}, vol.~48(6), pp.~1518-1569, 2002.

\bibitem{Pintelon01}
R.~Pintelon, J.~Schoukens,
``System Identification'',
\emph{IEEE Press}, Piscataway, NJ, 2001.

\bibitem{Rabiner89} L.R.~Rabiner:
``A tutorial on hidden Markov models and selected applications in speech recognition'',
{\em Proceedings of the IEEE}, vol.~77, pp.~257-286, 1989.

\bibitem{Schoenhuth07} A.~Sch\"onhuth, H.~Jaeger:
``Characterization of ergodic hidden Markov sources'',
{\em IEEE Transactions on Information Theory}, vol.~55, pp.~2107-2118, 2009.

\bibitem{Tzeng92} W.-G. Tzeng:
``A polynomial-time algorithm for the equivalence of probabilistic automata'',
{\em SIAM Journal of Computing}, vol.~21, pp.~216-227, 1992.

\end{thebibliography}
\end{document}